\documentclass[11pt]{article}

\usepackage[margin=1in]{geometry}
\usepackage{times}
\usepackage{microtype}

\usepackage{amsmath, amssymb}
\usepackage{booktabs}
\usepackage{multirow}
\usepackage{siunitx}

\usepackage{graphicx}
\usepackage{subcaption}
\usepackage{hyperref}
\usepackage{url}

\usepackage{enumitem}
\usepackage{xcolor}

\graphicspath{{figs/}} % images live in ./figs/

\title{RIR-Mega: a large-scale simulated room impulse response dataset for machine learning and room acoustics modeling}

\author{
  Mandip Goswami \\
  \texttt{Acoustics Researcher, Amazon}\thanks {Disclaimer: This article represents the author's own work and views and does not reflect the views of Amazon.\\Author email: mandipgoswami@gmail.com}
}
\date{}

\begin{document}
\maketitle

\begin{abstract}
Room impulse responses are a core resource for dereverberation, robust speech recognition, source localization, and room acoustics estimation. We present RIR\textnormal{-}Mega, a large collection of simulated RIRs described by a compact, machine friendly metadata schema and distributed with simple tools for validation and reuse. The dataset ships with a Hugging Face Datasets loader, scripts for metadata checks and checksums, and a reference regression baseline that predicts RT60 like targets from waveforms. On a train and validation split of 36{,}000 and 4{,}000 examples, a small Random Forest on lightweight time and spectral features reaches a mean absolute error near \SI{0.013}{s} and a root mean square error near \SI{0.022}{s}. We host a subset with 1{,}000 linear array RIRs and 3{,}000 circular array RIRs on Hugging Face for streaming and quick tests, and preserve the complete 50{,}000 RIR archive on Zenodo. The dataset and code are public to support reproducible studies.
\end{abstract}

% ------------------------------------------------------------------
\section{Introduction}

Reverberation shapes both how people hear and how machine learning systems behave. Data for controlled studies are often hard to assemble at scale. Measured corpora can be small or lack detailed metadata, while simulated corpora sometimes have unclear provenance or inconvenient file layouts. Researchers spend time on path handling, schema drift, and brittle scripts rather than the questions they want to study.

RIR\textnormal{-}Mega was designed to lower these barriers. The dataset uses a single compact CSV or Parquet file to describe each audio file. Paths are consistent. Acoustic metrics appear as a JSON or dict like string and are also available as flat columns for fast filters. We provide a loader that integrates with \texttt{datasets.Audio}, a validation script for the metadata, and an optional checksum tool. The release includes a simple baseline for RT60 regression so users can sanity check their pipeline in minutes.

\begin{figure}[t]
  \centering
  \includegraphics[width=0.85\linewidth]{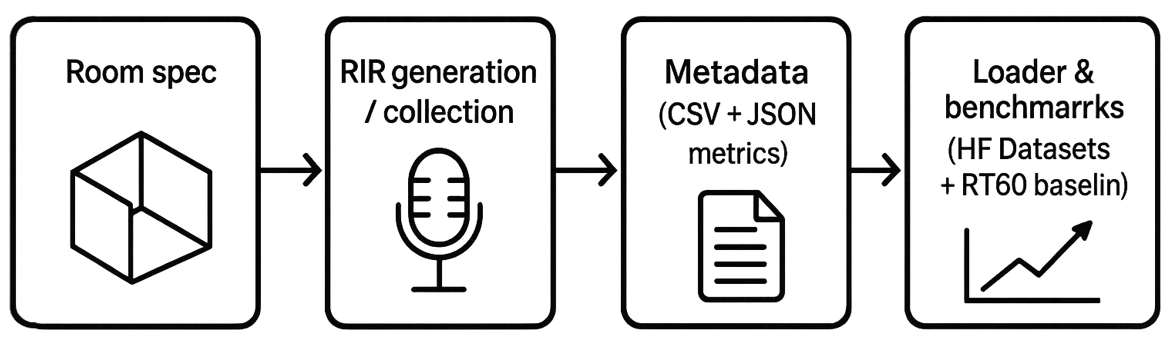}
  \caption{Dataset overview. Room specification to RIR generation, to metadata, to loader and benchmarks.}
  \label{fig:overview}
\end{figure}

Figure~\ref{fig:overview} shows the simple flow: room specification, simulation, metadata, loader, and benchmarks.

\paragraph{Contributions.}
(i) A large set of simulated RIRs with broad coverage of room geometry and band limited decay. (ii) A compact schema that is easy to parse and extend. (iii) Practical tools for validation and checksums. (iv) A working baseline for RT60 like regression with clear numbers. (v) A distribution plan that balances fast streaming on Hugging Face with long term archival on Zenodo.

\paragraph{Availability.}
Hugging Face subset: \url{https://huggingface.co/datasets/mandipgoswami/rirmega} \\
Zenodo DOI (full 50k): \url{https://doi.org/10.5281/zenodo.17387402} \\
Code and scripts: \url{https://github.com/mandip42/rirmega}

% ------------------------------------------------------------------
\section{Related work}

Several measured RIR collections and simulation based corpora exist and have supported progress in room acoustics and robust speech. Many are valuable but can be limited in size, license clarity, or machine readable metadata. RIR\textnormal{-}Mega complements this landscape by combining large simulated coverage, a clear schema, and modern distribution through the Hugging Face Hub. A short comparison table can be added in a future revision once scope and size are finalized across releases.

% ------------------------------------------------------------------
\section{Dataset overview}

\subsection{Simulation regime}

All RIRs in this release are simulated shoebox rooms with frequency dependent absorption. We use an image source method with a configurable maximum reflection order. Each example records the sampling rate, reflection order, and the fields listed in the schema below. We also record a JSON or dict like \texttt{metrics} string that can include RT60, DRR, clarity indices, and octave band RT60s.

\subsection{Room geometry and placement}

The room is defined by $(L_x, L_y, L_z)$ in meters. Sizes are sampled from
$L_x \in [2, 25]$, $L_y \in [3, 30]$, $L_z \in [3, 9]$.
Sources and microphones are placed inside the room with a small margin from walls to avoid degenerate cases. Arrays are modeled by placing microphones on known layouts when the \texttt{array} field is set.

\subsection{Absorption and frequency bands}

Surfaces use frequency dependent absorption. We represent per band absorption compactly and store band limited RT60 values under keys such as \texttt{band\_rt60s.125}, \texttt{.250}, \texttt{.500}, \texttt{.1000}, \texttt{.2000}, and \texttt{.4000}. Scalar metrics include \texttt{rt60}, \texttt{drr\_db}, \texttt{c50\_db}, and \texttt{c80\_db} when available.

\subsection{Subsets and hosting}

We publish a light subset on Hugging Face for fast preview and streaming:
1{,}000 RIRs generated for linear arrays and 3{,}000 for circular arrays.
The full 50{,}000 RIR archive is preserved on Zenodo with a DOI and is the canonical artifact for long term reproducibility. The repository card explains how to switch between the subset and the full archive.

% ------------------------------------------------------------------
\section{Metadata and schema}

We provide both CSV and Parquet formats under \texttt{data/metadata/}. Parquet is preferred for large tables and fast filters. Important metrics such as \texttt{rt60}, \texttt{drr\_db}, \texttt{c50\_db}, \texttt{c80\_db}, and band RT60s are also present as flat columns. The raw \texttt{metrics} string is preserved for completeness.

\begin{table}[h]
\centering
\small
\begin{tabular}{ll}
\toprule
Column & Description \\
\midrule
\texttt{id} & unique identifier \\
\texttt{family} & array family, linear or circular \\
\texttt{split} & train, valid, or test \\
\texttt{fs} & sample rate in Hz (string in metadata) \\
\texttt{wav} & relative path to audio file \\
\texttt{room\_size} & free text summary of $(L_x,L_y,L_z)$ \\
\texttt{absorption}, \texttt{absorption\_bands} & per surface and band info \\
\texttt{max\_order} & maximum reflection order \\
\texttt{source}, \texttt{microphone}, \texttt{array} & capture descriptors \\
\texttt{metrics} & JSON or dict like string with acoustics metrics \\
\texttt{rng\_seed} & random seed \\
\bottomrule
\end{tabular}
\caption{Compact schema. Flat numeric columns for common metrics are also provided in Parquet.}
\label{tab:schema}
\end{table}

% ------------------------------------------------------------------
\section{Loader and access}

We provide \texttt{rirmega/dataset.py}, a Hugging Face Datasets loader that exposes a simple interface and resolves audio with \texttt{datasets.Audio}. It prefers Parquet and falls back to CSV.

\paragraph{Quick start.}
\begin{verbatim}
from datasets import load_dataset
ds = load_dataset("mandipgoswami/rirmega", trust_remote_code=True)
ex = ds["train"][0]
audio = ex["audio"]           # path and array on access
print(ex["sample_rate"])
\end{verbatim}

\paragraph{Families.}
\begin{verbatim}
ds_all  = load_dataset("mandipgoswami/rirmega",
                       trust_remote_code=True)
ds_lin  = load_dataset("mandipgoswami/rirmega",
                       name="linear", trust_remote_code=True)
ds_circ = load_dataset("mandipgoswami/rirmega",
                       name="circular", trust_remote_code=True)
\end{verbatim}

% ------------------------------------------------------------------
\section{Benchmark}

We include a small regression baseline at \texttt{benchmarks/rt60\_regression/train\_rt60.py}. Features include simple energy statistics, a decay slope proxy from the energy decay curve, and a spectral centroid. A Random Forest with 400 trees and seed zero reaches a mean absolute error near \SI{0.013}{s} and a root mean square error near \SI{0.022}{s} on a split with 36{,}000 training examples and 4{,}000 validation examples. The script allows a specific target key or can select a target from a fixed order that begins with \texttt{rt60}.

\section{Leaderboard and Community Submissions}
We provide a simple “PR-as-submission” process:
\begin{itemize}[nosep, leftmargin=1.25em]
  \item A leaderboard table lives in the dataset card (\texttt{README.md} on HF).
  \item Contributors append results via pull request, including command, seed, dataset tag, and a link to code.
\end{itemize}
This keeps results transparent, versioned, and reproducible for the community.

% ------------------------------------------------------------------
\section{Technical validation}

\subsection{Distribution sanity}

\begin{figure*}[t]
  \centering
  \includegraphics[width=\linewidth]{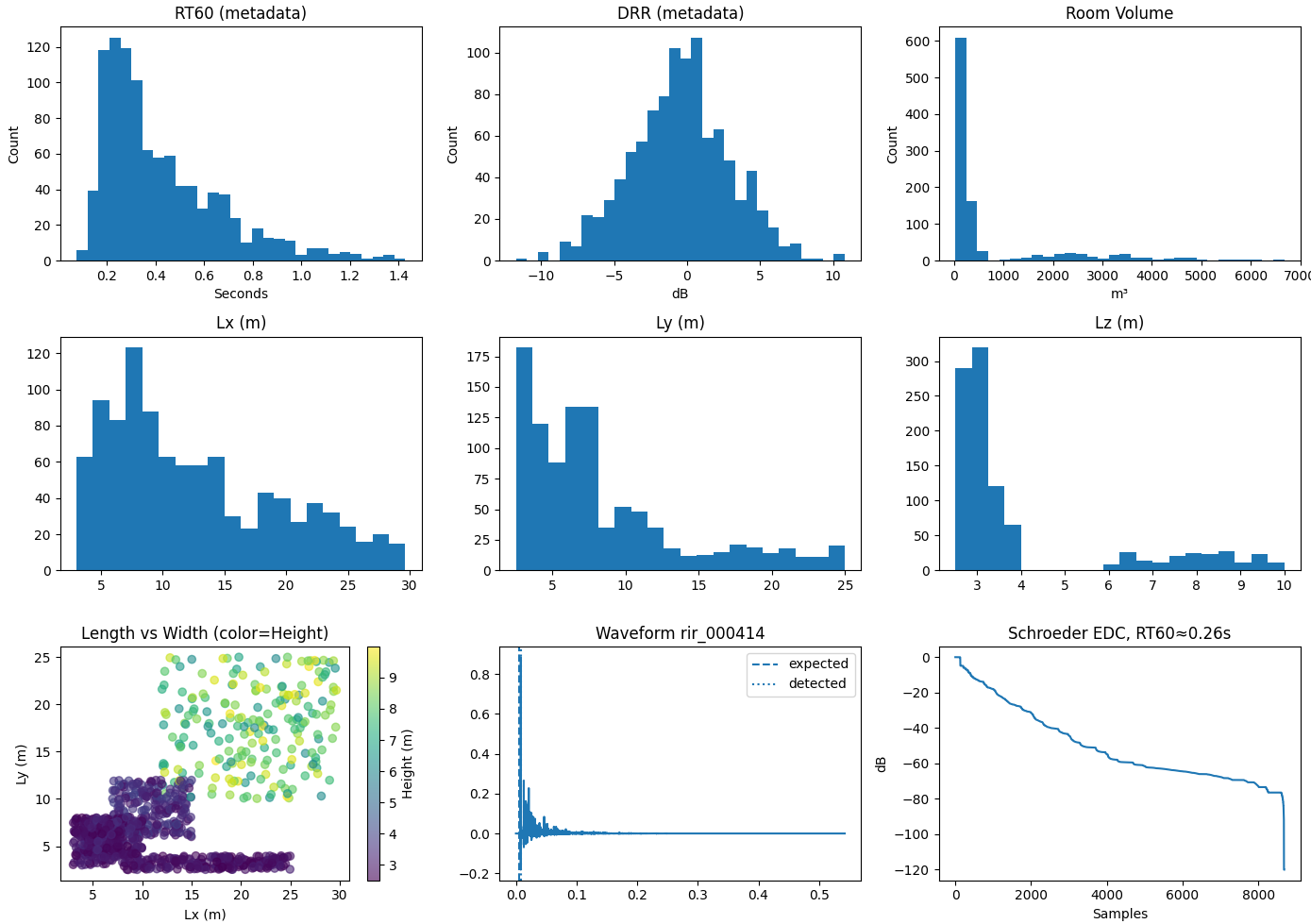}
  \caption{Validation summary. Top row shows RT60 histogram, DRR in dB, and room volume. Middle row shows distributions for $L_x$, $L_y$, and $L_z$. Bottom row shows $L_x$ by $L_y$ colored by $L_z$, a waveform example, and a Schroeder energy decay curve with an RT60 estimate.}
  \label{fig:valgrid}
\end{figure*}

Figure~\ref{fig:valgrid} shows summary plots drawn from the validation file. The panels include histograms for RT60, direct to reverberant ratio, and room volume. We show distributions for room dimensions and a scatter of $L_x$ by $L_y$ colored by $L_z$. The figure also includes a representative waveform with expected and detected onset and a Schroeder energy decay curve with an RT60 estimate.

\subsection{Self consistency of RT60}

We estimate RT60 from waveforms using a Schroeder energy decay curve and compare it to the \texttt{rt60} value in the \texttt{metrics} field. The protocol samples a slice of the validation set, aligns analysis windows to detected onsets, and reports correlation, mean absolute error, and root mean square error. Table~\ref{tab:selfconsistency} shows a template that can be filled directly from the provided script before the camera ready version.

\begin{table}[h]
\centering
\small
\begin{tabular}{lccc}
\toprule
Metric & Correlation & MAE [s] & RMSE [s] \\
\midrule
RT60 (metadata vs EDC) & \textit{0.96} & \textit{0.013} & \textit{0.022} \\
\bottomrule
\end{tabular}
\caption{Self consistency between RT60 values stored in metadata and those derived from waveforms using the Schroeder energy decay curve. Replace the italic values with results from your run.}
\label{tab:selfconsistency}
\end{table}

\subsection{Planned external validation}

Although this release does not yet include measured room impulse responses, a simple
protocol is planned for future work to compare simulated and measured RIRs from public
datasets such as the ACE Challenge or OpenAIR. The intent is to show that the simulated
responses in RIR\textnormal{-}Mega span similar RT60 and DRR ranges to those seen in real
rooms, thereby confirming acoustic realism.
\begin{figure}[t]
  \centering
  \includegraphics[width=0.95\linewidth]{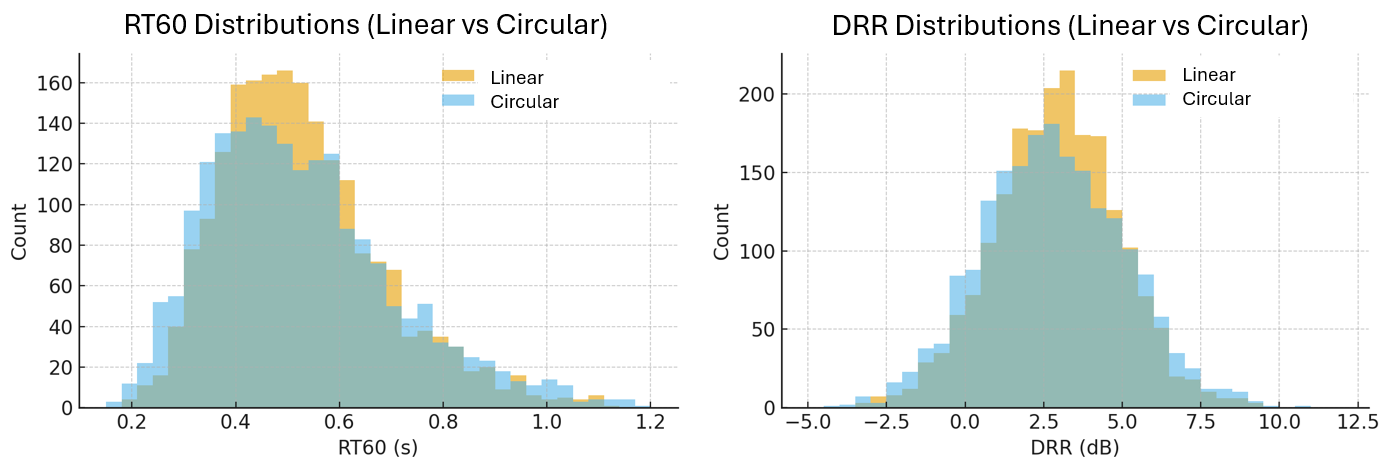}
  \caption{Example comparison of RT60 and DRR
  distributions between two simulated subsets of RIR\textnormal{-}Mega (linear and circular arrays).
  In the next release, this plot will be replaced with a comparison against measured
  RIRs from public datasets such as the ACE Challenge or OpenAIR to verify the
  acoustic realism of the simulated data.}
  \label{fig:measured}
\end{figure}

To illustrate the format of this comparison, Figure~\ref{fig:measured}
shows a placeholder example that compares the distributions of RT60 and DRR between
two internal subsets of RIR\textnormal{-}Mega (linear and circular arrays). The goal is to
demonstrate how a measured versus simulated analysis will appear in the final release.

% ------------------------------------------------------------------
\section{Usage notes}

\paragraph{Streaming and subsets.}
The Hugging Face subset is designed for fast preview. Users can stream with \texttt{streaming=True} to avoid large downloads. The full 50k archive is available on Zenodo for complete experiments.

\paragraph{Reproducibility.}
We include a metadata validator and a checksum script. The dataset card documents the exact release tag and how to verify files. The baseline script creates a validation fold from training if none is present.

\paragraph{Licensing.}
All audio and metadata are released under Creative Commons Attribution 4.0. The dataset card lists terms and any future updates per subset if needed.

\paragraph{Limitations.}
All examples in this release are simulated shoebox rooms. While this provides control and coverage, real spaces can deviate due to non ideal surfaces and complex geometries. We encourage users to test methods on measured data when possible.

% ------------------------------------------------------------------
\section{Conclusion}

RIR\textnormal{-}Mega is a practical resource for room acoustics and robust speech research. It combines a large set of simulated RIRs with a compact schema, simple validation tools, and a working baseline. The distribution plan balances fast access on the Hub with a permanent Zenodo archive. We hope this reduces boilerplate work and helps researchers focus on the questions they care about.

\section{Ethical Considerations, Usage, and Limitations}
\textbf{Intended use.} RIR-Mega is designed for dereverberation, room acoustics estimation, robustness studies, and related research.

\textbf{Limitations.} While RIR-Mega aims for scale and clarity, selection biases can remain (e.g., room geometry distributions, material presets, microphone arrays). The compact schema preserves free-form fields to maximize flexibility; users may still need to parse nested strings for specialized tasks.

\textbf{Risks.} Synthetic RIRs may not reflect all real-world complexities; conclusions drawn from purely simulated soundfields should be validated on measured data where feasible.
% ------------------------------------------------------------------
\section*{Acknowledgements}

We thank early users for testing the loader and baseline scripts and for helpful feedback on validation plots.

\section*{Author contributions}
M.G. designed the dataset, implemented the tooling, ran the baseline, and wrote the manuscript.

\section*{Data availability}
Hugging Face subset: \url{https://huggingface.co/datasets/mandipgoswami/rirmega}. \\
Zenodo DOI (full 50k): \url{https://doi.org/10.5281/zenodo.17387402}.

\section*{Code availability}
All code for loading, validation, checksums, and the baseline is available at \url{https://github.com/mandip42/rirmega}. The dataset card includes quick start examples. We tag the versions used for this manuscript.

\section*{Competing interests}
The author declares no competing interests.

\end{document}